# Fuzzy finite element solution of uncertain neutron diffusion equation for imprecisely defined homogeneous triangular bare reactor

S. Nayak and S. Chakraverty*
Department of Mathematics, National Institute of Technology, Rourkela, Odisha -769008, India

**Abstract:** Scattering of neutron collision inside a reactor depends upon geometry of the reactor, diffusion coefficient and absorption coefficient etc. In general these parameters are not crisp and hence we may get uncertain neutron diffusion equation. In this paper we have investigated the above problem for a bare triangular homogeneous reactor. Here the uncertain governing differential equation is modelled by a modified fuzzy finite element method using newly proposed interval arithmetic. Obtained eigenvalues by the proposed method are studied in detail. Further the eigenvalues are compared with the classical finite element method in special cases and various uncertain results have been discussed.

## 1. Introduction

In a nuclear reactor neutrons suffer scattering collisions with the nuclei assumed to be initially stationary and as a result a typical neutron trajectory consists of a number of short straight path elements. These are the scattering free paths. The average of these is the mean free paths. It may be noted that the path of neutron after a scattering collision is not known exactly, so there may involve some uncertainty. These uncertainties may occur due to incomplete data, impreciseness, vagueness, experimental error and different operating conditions influenced by the system. Different authors proposed various methods to handle

*corresponding author
E-mail address: sne_chak@yahoo.com (S. Chakraverty), sukantgacr@gmail.com (S. Nayak)

these uncertainties. They have used probabilistic or statistical method as a tool to handle the uncertain parameters. In this Context, Monte Carlo method is an alternate way which is based on the statistical simulation of the random numbers generated on the basis of a specific sampling distribution. Monte Carlo methods have been used by many authors to solve neutron diffusion equation with variable parameters. As such, Nagaya et al. [7] implemented Monte Carlo method to estimate the effective delayed neutron fraction $\beta_{eff}$ . Further, Nagaya and Mori [8] proposed a new method to estimate the effective delayed neutron fraction $\beta_{eff}$ in Monte Carlo calculations. In the above paper, the eigenvalue method is jointly used with the differential operator and correlated sampling techniques. Shi and Petrovic [10] used Monte Carlo method to solve one-dimensional two-group problems. Sjenitzer and Hoogenboom [11] gave an analytical procedure to compute the variance of the neutron flux in a simple model of a fixed-source. Recently, Yamamoto [12] investigated the neutron leakage effect specified by buckling to generate group constants for use in reactor core designs using Monte Carlo method.

In the above procedure we need a good number of observed data or experimental results to analyse the problem. However, it may not be possible always to get a large number of data. As regards; Zadeh [13] proposed an alternate idea viz. fuzzy approach to handle uncertain and imprecise variables. Accordingly, we may use interval or fuzzy parameters to take care of the uncertainty. In general traditional interval/fuzzy arithmetic are complicated to investigate the problem. Here we have proposed a new procedure of fuzzy arithmetic to overcome such difficulty. The idea as proposed by Chakraverty and Nayak [4, 9] has been extended. Here our aim is to investigate neutron diffusion equation. In this respect, Biswas et al. [3] have given a method of generating stiffness matrices for the solution of multi group diffusion equation by natural coordinate system. Azekura [1] has also proposed a new representation of finite element solution technique for neutron diffusion equations. The author has applied the technique to two types of one-group neutron diffusion equations to test its accuracy. It reveals from the above literature that the neutron diffusion equations are solved using finite element method in presence of crisp parameters only.

But the presence of uncertain parameters makes the system uncertain and we get uncertain governing differential equations. To the best of our knowledge, no study has been done for fuzzy/interval uncertain neutron diffusion equation. In this context, uncertain fuzzy parameters are considered to solve heat conduction problems using finite element method and we call it as Fuzzy Finite Element Method (FFEM). Recently Bart et al. [2] solved the uncertain solution of heat conduction problem. In this paper authors gave a good comparison between response surface method and other methods. Recently, Chakraverty and Nayak [5] also solved the interval/fuzzy distribution of effective multiplication factors and eigenvalues for a bare square homogeneous reactor.

In view of the above, here we present a modified form of fuzzy finite element method. The involved fuzzy numbers are changed into intervals through $\alpha$-cut. Then the intervals are transformed into crisp form by using some transformations. Crisp representations of intervals are defined by symbolic parameterization. Traditional interval arithmetic is modified using the crisp representation of intervals. The proposed interval arithmetic is extended for fuzzy numbers and the developed fuzzy arithmetic is used as a tool for uncertain fuzzy finite element method. Consequently the above method is used to solve one group neutron diffusion equation for triangular bare reactor and the critical eigenvalues are studied in detail. Finally some important conclusions of the proposed methods are encrypted and it is found that this method is simpler and efficient to handle. Hence it may be used as a tool to solve different types of neutron diffusion problems for various types of nuclear reactors.

## 2. Interval and Fuzzy Arithmetic

The uncertain values occurred in practical cases (such as experimental data, impreciseness and partial or imperfect knowledge) may be handled by taking the uncertainty as interval or fuzzy sense. So to compute these uncertainties we need interval/fuzzy arithmetic. Let us consider the uncertain values in interval form and the same may be written in the following way.

$$[\underline{x}, \overline{x}] = \{x | x \in R, \underline{x} \le x \le \overline{x}\}$$

where $\underline{x}$ and $\overline{x}$ are lower and upper values of the interval respectively. Let us consider $m = \frac{\underline{x} + \overline{x}}{2}$ and $w = \overline{x} - \underline{x}$ are centre and width of the interval $[\underline{x}, \overline{x}]$ respectively.

Let us assume that $[\underline{x}, \overline{x}]$ and $[\underline{y}, \overline{y}]$ be two intervals then we have standard interval arithmetic as,

1. $[\underline{x}, \overline{x}] + [\underline{y}, \overline{y}] = [\underline{x} + \underline{y}, \overline{x} + \overline{y}]$
2. $[\underline{x}, \overline{x}] - [\underline{y}, \overline{y}] = [\underline{x} - \overline{y}, \overline{x} - \underline{y}]$
3. $[\underline{x}, \overline{x}] \times [\underline{y}, \overline{y}] = [\min\{\underline{x}\underline{y}, \underline{x}\overline{y}, \overline{x}\underline{y}, \overline{x}\overline{y}\}, \max\{\underline{x}\underline{y}, \underline{x}\overline{y}, \overline{x}\underline{y}, \overline{x}\overline{y}\}]$
4. $[\underline{x}, \overline{x}] \div [\underline{y}, \overline{y}] = [\min\{\underline{x} \div \underline{y}, \underline{x} \div \overline{y}, \overline{x} \div \underline{y}, \overline{x} \div \overline{y}\}, \max\{\underline{x} \div \underline{y}, \underline{x} \div \overline{y}, \overline{x} \div \underline{y}, \overline{x} \div \overline{y}\}]$

We may extend this concept into various fuzzy numbers viz. triangular and trapezoidal fuzzy numbers etc. Any arbitrary fuzzy number may be defined in terms of interval involving left and right continuous linear functions. Fuzzy numbers may be represented as an ordered pair form $[\underline{f}(\alpha), \overline{f}(\alpha)]$, $0 \le \alpha \le 1$ where $\underline{f}(\alpha)$ and $\overline{f}(\alpha)$ are left and right monotonic increasing and decreasing functions over [0, 1] respectively.

Let us consider two fuzzy numbers $x = [\underline{x}(\alpha), \overline{x}(\alpha)]$ and $y = [\underline{y}(\alpha), \overline{y}(\alpha)]$ and a scalar $k$ then

i. $x = y$ if and only if $\underline{x}(\alpha) = \underline{y}(\alpha)$ and $\overline{x}(\alpha) = \overline{y}(\alpha)$.

ii. $x + y = [\underline{x}(\alpha) + \underline{y}(\alpha), \overline{x}(\alpha) + \overline{y}(\alpha)]$.

iii. $kx = \begin{cases} [k\underline{x}(\alpha), k\overline{x}(\alpha)], & k \ge 0, \\ [k\overline{x}(\alpha), k\underline{x}(\alpha)], & k < 0. \end{cases}$

**Definition 2.1**

Above interval arithmetic for real interval are defined here as follows [4]

1. $[\underline{x},\overline{x}]+[\underline{y},\overline{y}]$=[min{ $\lim\limits_{n\to\infty} l_1+\lim\limits_{n\to\infty} l_2, \lim\limits_{n\to 1} l_1+\lim\limits_{n\to 1} l_2$ }, max{ $\lim\limits_{n\to\infty} l_1+\lim\limits_{n\to\infty} l_2, \lim\limits_{n\to 1} l_1+\lim\limits_{n\to 1} l_2$ }]

2. $[\underline{x},\overline{x}]-[\underline{y},\overline{y}]$=[min{ $\lim\limits_{n\to\infty} l_1-\lim\limits_{n\to 1} l_2, \lim\limits_{n\to 1} l_1-\lim\limits_{n\to\infty} l_2$ }, max{ $\lim\limits_{n\to\infty} l_1-\lim\limits_{n\to 1} l_2, \lim\limits_{n\to 1} l_1-\lim\limits_{n\to\infty} l_2$ }]

3. $[\underline{x},\overline{x}]\times[\underline{y},\overline{y}]$=[min{ $\lim\limits_{n\to\infty} l_1\times\lim\limits_{n\to\infty} l_2, \lim\limits_{n\to 1} l_1\times\lim\limits_{n\to 1} l_2$ }, max{ $\lim\limits_{n\to\infty} l_1\times\lim\limits_{n\to\infty} l_2, \lim\limits_{n\to 1} l_1\times\lim\limits_{n\to 1} l_2$ }]

4. $[\underline{x},\overline{x}]\div[\underline{y},\overline{y}]$=[min{ $\lim\limits_{n\to\infty} l_1\div\lim\limits_{n\to 1} l_2, \lim\limits_{n\to 1} l_1\div\lim\limits_{n\to\infty} l_2$ }, max{ $\lim\limits_{n\to\infty} l_1\div\lim\limits_{n\to 1} l_2, \lim\limits_{n\to 1} l_1\div\lim\limits_{n\to\infty} l_2$ }]

where for an arbitrary interval $[\underline{a},\overline{a}]=\left\{\underline{a}+\frac{w}{n}=l \mid \underline{a}\le l\le\overline{a}, n\in[1,\infty)\right\}$ and $w=\overline{a}-\underline{a}$ is the width of the interval.

**Definition 2.2**

A fuzzy number $\tilde{A}=[a^L,a^N,a^R]$ is said to be triangular fuzzy number (Figure 1) when the membership function is given by

$$\mu_{\tilde{A}}(x)=\begin{cases} 0, & x\le a^L; \\ \dfrac{x-a^L}{a^N-a^L}, & a^L\le x\le a^N; \\ \dfrac{a^R-x}{a^R-a^N}, & a^N\le x\le a^R; \\ 0, & x\ge a^R. \end{cases}$$

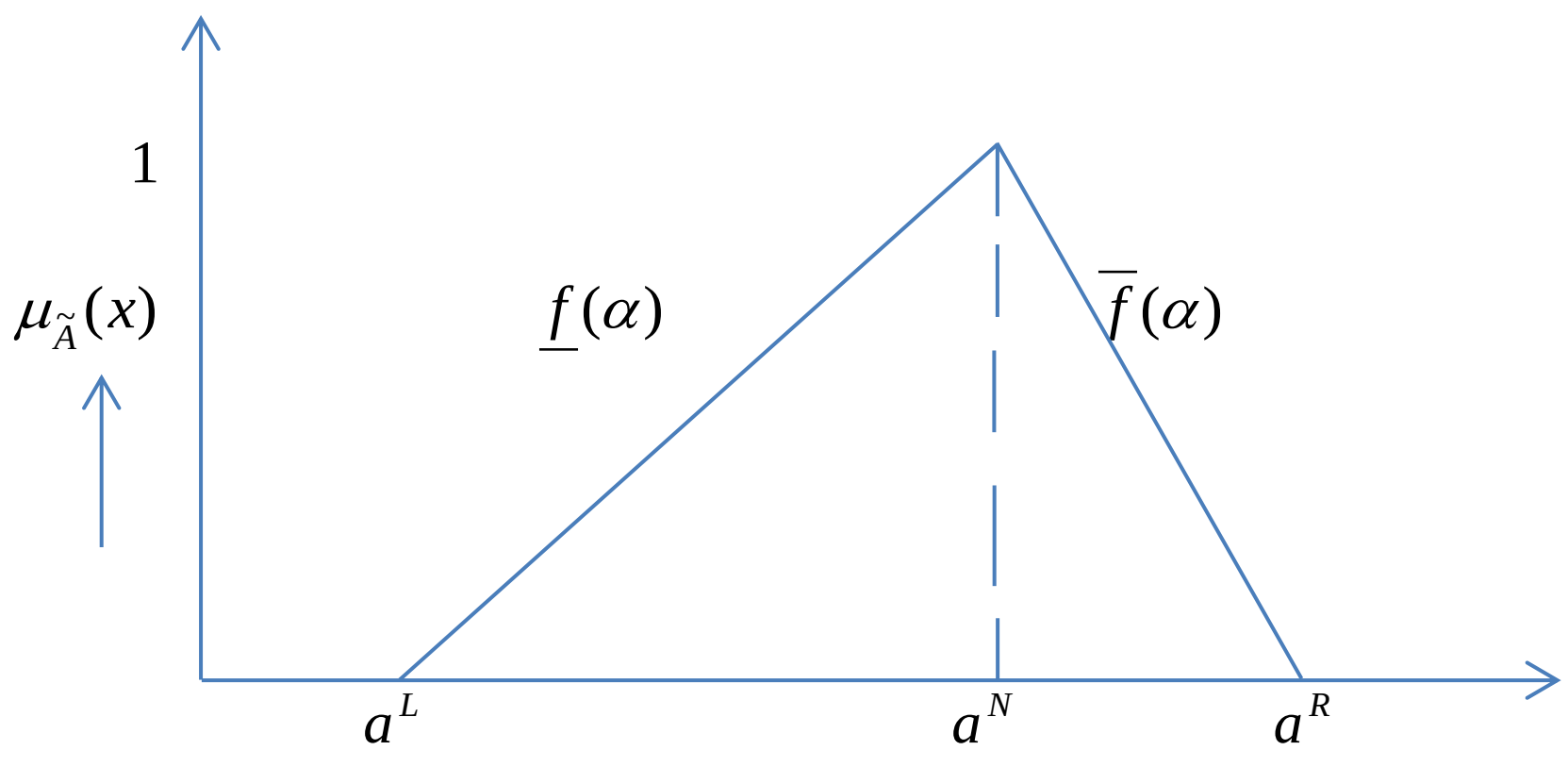

Figure 1 Triangular Fuzzy Number (TFN)

**Definition 2.4**

The triangular fuzzy number $\tilde{A} = [a^L, a^N, a^R]$ may be transformed into interval form by using $\alpha -$cut in the following form

$$\tilde{A} = [a^L, a^N, a^R] = [a^L + (a^N - a^L)\alpha,\ a^R - (a^R - a^N)\alpha],$$

**Definition 2.5**

If the fuzzy numbers are taken in interval form then using definition 2.1, the arithmetic rules may be defined as

1. $[\underline{x}(\alpha), \bar{x}(\alpha)] + [\underline{y}(\alpha), \bar{y}(\alpha)]$

   $= [\min\{\lim_{n\to\infty} m_1 + \lim_{n\to\infty} m_2, \lim_{n\to 1} m_1 + \lim_{n\to 1} m_2\}, \max\{\lim_{n\to\infty} m_1 + \lim_{n\to\infty} m_2, \lim_{n\to 1} m_1 + \lim_{n\to 1} m_2\}]$

2. $[\underline{x}(\alpha), \bar{x}(\alpha)] - [\underline{y}(\alpha), \bar{y}(\alpha)]$

   $= [\min\{\lim_{n\to\infty} m_1 - \lim_{n\to 1} m_2, \lim_{n\to 1} m_1 - \lim_{n\to\infty} m_2\}, \max\{\lim_{n\to\infty} m_1 - \lim_{n\to 1} m_2, \lim_{n\to 1} m_1 - \lim_{n\to\infty} m_2\}]$

3. $[\underline{x}(\alpha), \bar{x}(\alpha)] \times [\underline{y}(\alpha), \bar{y}(\alpha)]$

   $= [\min\{\lim_{n\to\infty} m_1 \times \lim_{n\to\infty} m_2, \lim_{n\to 1} m_1 \times \lim_{n\to 1} m_2\}, \max\{\lim_{n\to\infty} m_1 \times \lim_{n\to\infty} m_2, \lim_{n\to 1} m_1 \times \lim_{n\to 1} m_2\}]$

4. $[\underline{x}(\alpha), \bar{x}(\alpha)] \div [\underline{y}(\alpha), \bar{y}(\alpha)]$

   $= [\min\{\lim_{n\to\infty} m_1 \div \lim_{n\to 1} m_2, \lim_{n\to 1} m_1 \div \lim_{n\to\infty} m_2\}, \max\{\lim_{n\to\infty} m_1 \div \lim_{n\to 1} m_2, \lim_{n\to 1} m_1 \div \lim_{n\to\infty} m_2\}]$

where for any arbitrary interval

$$[\underline{f}(\alpha), \overline{f}(\alpha)] = \left\{ \underline{f}(\alpha) + \frac{\overline{f}(\alpha) - \underline{f}(\alpha)}{n} = m \,\middle|\, \underline{f}(\alpha) \le m \le \overline{f}(\alpha), n \in [1, \infty) \right\}.$$

Various problems of science and engineering always involve some form of uncertainty. This uncertainty here we consider in term of fuzzy or interval. While solving different problems by traditional FEM we usually consider the associated parameters and formulations in crisp form. But to handle the uncertainty while using FEM, we must formulate the FEM in uncertainty form (fuzzy or interval) using the fuzzy or interval computation. In this paper we formulate the problem first as traditional FEM for the sake of completeness. Then the problem has been formulated considering the uncertainty. Henceforth the main aim of this paper is to solve the uncertain one group neutron diffusion equation for triangular bare reactor by an alternate non-probabilistic method presented and we call it FFEM or IFEM.

**3. Formulation of the problem**

As it is known that the principle of neutron conservation can be expressed in a simple form for a system of mono energetic neutrons. Multi group equations can be analysed by considering the series of one group equations.

The standard functional for corresponding one group diffusion equation may be written as

$$I(\phi)=\frac{1}{2}\iint_R\left[D\left(\frac{\partial\phi}{\partial x}\right)^2+D\left(\frac{\partial\phi}{\partial y}\right)^2+\sigma\phi^2-2S\phi\right]dxdy \quad (1)$$

where $\phi$ is a constant over a partial/total portion of the periphery, $D$ is the diffusion coefficient, $\sigma$ is the absorption coefficient and $S$ is the source term.

If we apply traditional FEM to handle the problem the corresponding domain of the problem is divided into number of subdomain and each of them is called element. For each element we may find the functional and similarly for the entire domain the functional may be found out by summing each functional element wise. The above procedure may be written in the following way.

First the domain $R$ may be represented as

$$R=\sum_{e=1}^{n}R^e \quad (2)$$

and the functional $I(\phi)$ is defined as

$$I(\phi)=\sum_{e=1}^{n}I^e(\phi) \quad (3)$$

where $n$ is the total number of elements and $I^e(\phi)$ denotes the contribution of element $e$ to the functional $I(\phi)$. Now the equation (1) for each elemental functional may be written as

$$I^e(\phi) = \frac{1}{2}\iint_R \left[ D^e\left(\frac{\partial\phi^e}{\partial x}\right)^2 + D^e\left(\frac{\partial\phi^e}{\partial y}\right)^2 + \sigma^e\phi^2 - 2S^e\phi^e \right] dxdy \tag{4}$$

For each element $e$ the scalar flux $\phi^e$ is approximated by a piecewise interpolation polynomial. Depending on the interpolation polynomial, stiffness matrices are obtained by minimizing the elemental functional $I^e(\phi)$. The stiffness matrices are assembled and finally we get the algebraic form which is represented as

$$[K]\{\phi\} = \{Q\} \tag{5}$$

where $[K]$ is the assembled stiffness matrix corresponding to leakage and absorption terms and $\{Q\}$ is the assembled force vector for the source term.

In general when neutrons undergo scattering, the neutron transport equation involves uncertainty. The uncertainty occurs due to the imprecise value of operating parameters viz. geometry, diffusion and absorption coefficients etc. Here these uncertain parameters are taken as fuzzy. To investigate the uncertain spectrum of neutron flux distribution we formulate fuzzy finite element method with linear triangular fuzzy element discretising the domain.

Let us consider that the coordinates of linear triangular elements are in fuzzy and hence we may write

$$\begin{aligned} \tilde{x} &= L_1\tilde{x}_1 + L_2\tilde{x}_2 + L_3\tilde{x}_3; \\ \tilde{y} &= L_1\tilde{y}_1 + L_2\tilde{y}_2 + L_3\tilde{y}_3; \\ \tilde{L} &= \tilde{L}_1 + \tilde{L}_2 + \tilde{L}_3; \end{aligned} \tag{6}$$

where $\tilde{L}_i$ $(i = 1, 2, 3)$ are nondimensionalized coordinates.

The above Eq. (6) in matrix form is represented in the following way

$$\begin{bmatrix} 1 & 1 & 1 \\ \tilde{x}_1 & \tilde{x}_2 & \tilde{x}_3 \\ \tilde{y}_1 & \tilde{y}_2 & \tilde{y}_3 \end{bmatrix} \begin{Bmatrix} \tilde{L}_1 \\ \tilde{L}_2 \\ \tilde{L}_3 \end{Bmatrix} = \begin{Bmatrix} 1 \\ \tilde{x} \\ \tilde{y} \end{Bmatrix}$$

$$\Rightarrow \begin{Bmatrix} \tilde{L}_1 \\ \tilde{L}_2 \\ \tilde{L}_3 \end{Bmatrix} = \begin{bmatrix} 1 & 1 & 1 \\ \tilde{x}_1 & \tilde{x}_2 & \tilde{x}_3 \\ \tilde{y}_1 & \tilde{y}_2 & \tilde{y}_3 \end{bmatrix}^{-1} \begin{Bmatrix} 1 \\ \tilde{x} \\ \tilde{y} \end{Bmatrix}$$

$$\Rightarrow \begin{Bmatrix} \tilde{L}_1 \\ \tilde{L}_2 \\ \tilde{L}_3 \end{Bmatrix} = \begin{bmatrix} \tilde{x}_2\tilde{y}_3 - \tilde{x}_3\tilde{y}_2 & \tilde{y}_2 - \tilde{y}_3 & \tilde{x}_3 - \tilde{x}_2 \\ \tilde{x}_3\tilde{y}_1 - \tilde{x}_1\tilde{y}_3 & \tilde{y}_3 - \tilde{y}_1 & \tilde{x}_1 - \tilde{x}_3 \\ \tilde{x}_1\tilde{y}_2 - \tilde{x}_2\tilde{y}_1 & \tilde{y}_1 - \tilde{y}_2 & \tilde{x}_2 - \tilde{x}_1 \end{bmatrix} \begin{Bmatrix} 1 \\ \tilde{x} \\ \tilde{y} \end{Bmatrix}$$

where area of the fuzzy triangle is $\tilde{\Delta} = \frac{1}{2}\begin{bmatrix} 1 & 1 & 1 \\ \tilde{x}_1 & \tilde{x}_2 & \tilde{x}_3 \\ \tilde{y}_1 & \tilde{y}_2 & \tilde{y}_3 \end{bmatrix}$.

We now denote

$$\tilde{a}_1 = \tilde{x}_3 - \tilde{x}_2, \tilde{a}_2 = \tilde{x}_1 - \tilde{x}_3, \tilde{a}_3 = \tilde{x}_2 - \tilde{x}_1;$$
$$\tilde{b}_1 = \tilde{y}_2 - \tilde{y}_3, \tilde{b}_2 = \tilde{y}_3 - \tilde{y}_1, \tilde{b}_3 = \tilde{y}_1 - \tilde{y}_2;$$
$$\tilde{c}_1 = \tilde{x}_2\tilde{y}_3 - \tilde{x}_3\tilde{y}_2, \tilde{c}_2 = \tilde{x}_3\tilde{y}_1 - \tilde{x}_1\tilde{y}_3, \tilde{c}_3 = \tilde{x}_1\tilde{y}_2 - \tilde{x}_2\tilde{y}_1.$$

If $\tilde{\phi}$ is the flux distribution then it may be written as

$$\tilde{\phi} = \tilde{L}_1\tilde{\phi}_1 + \tilde{L}_2\tilde{\phi}_2 + \tilde{L}_3\tilde{\phi}_3. \tag{7}$$

The differentiation and integration formulae are then given by,

$$\frac{\partial}{\partial \tilde{x}} = \sum_{i=1}^{3} \frac{\tilde{b}_i}{2\tilde{\Delta}} \frac{\partial}{\partial \tilde{L}_i}, \quad \frac{\partial}{\partial \tilde{y}} = \sum_{i=1}^{3} \frac{\tilde{a}_i}{2\tilde{\Delta}} \frac{\partial}{\partial \tilde{L}_i} \text{ and } \iint_R \tilde{L}_1^p \tilde{L}_2^q \tilde{L}_3^r \, d\tilde{\Delta} = \frac{p!q!r!}{(p+q+r+2)!}(2\tilde{\Delta}).$$

Hence

$$\frac{\partial \tilde{\phi}^{(e)}}{\partial \tilde{x}} = \frac{1}{2\tilde{\Delta}}\left\{\tilde{b}_1\tilde{\phi}_1 + \tilde{b}_2\tilde{\phi}_2 + \tilde{b}_3\tilde{\phi}_3\right\} = \begin{bmatrix} \frac{\tilde{b}_1}{2\tilde{\Delta}} & \frac{\tilde{b}_2}{2\tilde{\Delta}} & \frac{\tilde{b}_3}{2\tilde{\Delta}} \end{bmatrix} \begin{Bmatrix} \tilde{\phi}_1 \\ \tilde{\phi}_2 \\ \tilde{\phi}_3 \end{Bmatrix}$$

Similarly,

$$\frac{\partial \tilde{\phi}^{(e)}}{\partial \tilde{y}} = \frac{1}{2\tilde{\Delta}}\left\{\tilde{a}_1\tilde{\phi}_1 + \tilde{a}_2\tilde{\phi}_2 + \tilde{a}_3\tilde{\phi}_3\right\} = \begin{bmatrix} \frac{\tilde{a}_1}{2\tilde{\Delta}} & \frac{\tilde{a}_2}{2\tilde{\Delta}} & \frac{\tilde{a}_3}{2\tilde{\Delta}} \end{bmatrix} \begin{Bmatrix} \tilde{\phi}_1 \\ \tilde{\phi}_2 \\ \tilde{\phi}_3 \end{Bmatrix}.$$

Using above formulation one may get the leakage and absorption stiffness matrices. Accordingly, corresponding stiffness matrices of each element for leakage and absorption term is given by,

$$\left[\tilde{K}_1\right] = \frac{\tilde{D}^{(e)}}{4\tilde{\Delta}} \begin{bmatrix} \tilde{a}_1^{\,2} + \tilde{b}_1^{\,2} & \tilde{a}_1\tilde{a}_2 + \tilde{b}_1\tilde{b}_2 & \tilde{a}_1\tilde{a}_3 + \tilde{b}_1\tilde{b}_3 \\ \tilde{a}_1\tilde{a}_2 + \tilde{b}_1\tilde{b}_2 & \tilde{a}_2^{\,2} + \tilde{b}_2^{\,2} & \tilde{a}_2\tilde{a}_3 + \tilde{b}_2\tilde{b}_3 \\ \tilde{a}_1\tilde{a}_3 + \tilde{b}_1\tilde{b}_3 & \tilde{a}_2\tilde{a}_3 + \tilde{b}_2\tilde{b}_3 & \tilde{a}_3^{\,2} + \tilde{b}_3^{\,2} \end{bmatrix} \text{ and } \left[\tilde{K}_2\right] = \frac{\tilde{\sigma}^{(e)}\tilde{\Delta}}{12} \begin{bmatrix} 2 & 1 & 1 \\ 1 & 2 & 1 \\ 1 & 1 & 2 \end{bmatrix} \text{ respectively.}$$

The source vector $\left\{\tilde{f}\right\}$ for each element may be written as

$$\left\{\tilde{f}\right\} = \frac{\tilde{S}^{(e)}\tilde{\Delta}}{3} \begin{Bmatrix} 1 \\ 1 \\ 1 \end{Bmatrix}.$$

The above discussed fuzzy arithmetic in terms of $\alpha$-cut has been used here for finite element method. Using new transformed fuzzy finite element method the uncertain fuzzy parameters are handled. The schematic diagram is presented below in Figure 2, which gives the overall idea to encrypt the process of modified fuzzy finite element method. It involves three steps such as input, output and hidden layer. The uncertain parameters involved in problems are taken as fuzzy. These parameters are accumulated through a process viz. modified fuzzy finite element method which plays the role of hidden layer. Using the fuzzy input parameters, hidden layer gives corresponding type of fuzzy solutions. Here in Figure 2 we have considered triangular fuzzy numbers as input parameters. Alpha level representation of two fuzzy sets $\tilde{X}$ and $\tilde{Y}$ with their triangular membership functions for fuzzy arithmetic operation [7] is shown in Figure 2. Deterministic value is obtained for $\alpha_4$-level of fuzzy sets whereas for $\alpha_1$, $\alpha_2$ and $\alpha_3$ level we get different interval values. If we consider the alpha value as zero then deterministic interval lies on X- axis. The output may be generated by considering all possible combinations of the alpha level.

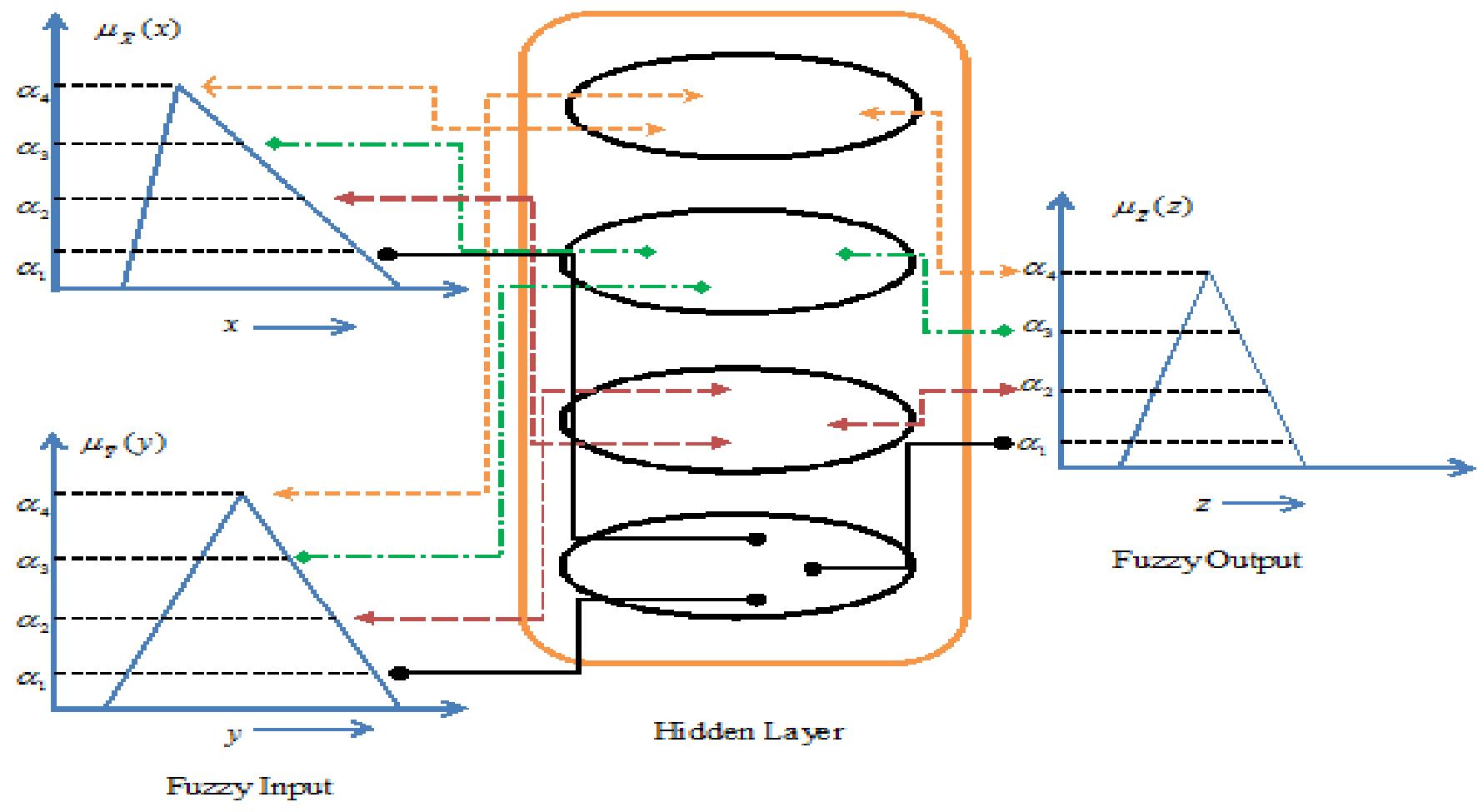

Figure 2 Model diagram of the modified fuzzy finite element procedure

# 4. Numerical Example

The governing differential equation for the bare homogeneous reactor [12] is as follows

$$D\nabla^2\phi + S = \Sigma_a\phi \tag{8}$$

We have considered a triangular (equilateral) bare reactor having each side of 4 units and it is discretized into triangular element as given in Figure 3.

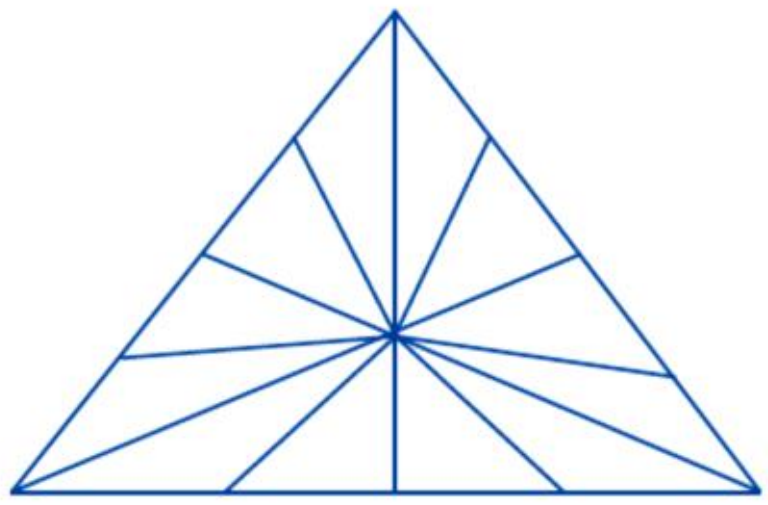

Figure 3 Triangular element discretization of triangular plate.

Fuzzy parameters are taken for diffusion and absorption coefficients which are presented in Table 1.

Table 1. Triangular fuzzy numbers for uncertain parameters

| Parameters | Crisp value | TFN |
|---|---|---|
| Diffusion coefficient | 1 | $[0.5+0.5\alpha, 1.5-0.5\alpha]$ |
| Absorption coefficient | 1 | $[0.5+0.5\alpha, 1.5-0.5\alpha]$ |

Initially the governing one group neutron diffusion equation is solved by considering only crisp parameters and then the proposed method for the modelled uncertain one group neutron diffusion equation is solved. Eigenvalues for both the crisp and fuzzy parameters are obtained and the values are depicted in Table 2 for different number of elements in the FEM and FFEM discretization.

Table 2. Crisp and triangular fuzzy eigenvalues for triangular plate

| Number of elements | Crisp eigenvalues | Triangular fuzzy eigenvalues |
|---|---|---|
| 6 | 0.6425 | [0.6377, 0.6425, 0.647] |
| 12 | 0.6264 | [0.6236, 0.6264, 0.6297] |
| 24 | 0.526 | [0.5251, 0.526, 0.527] |
| 48 | 0.5083 | [0.508, 0.5083, 0.5087] |
| 96 | 0.5034 | [0.5032, 0.5034, 0.5036] |
| 192 | 0.5015 | [0.5015, 0.5015, 0.5016] |
| 384 | 0.5007 | [0.5007, 0.5007, 0.5008] |
| 1536 | 0.5002 | [0.5002, 0.5002, 0.5002] |

For better visualization of the obtained results, eigenvalues for different number of discretizations of the domain are plotted and shown in Figures 4 to 11.

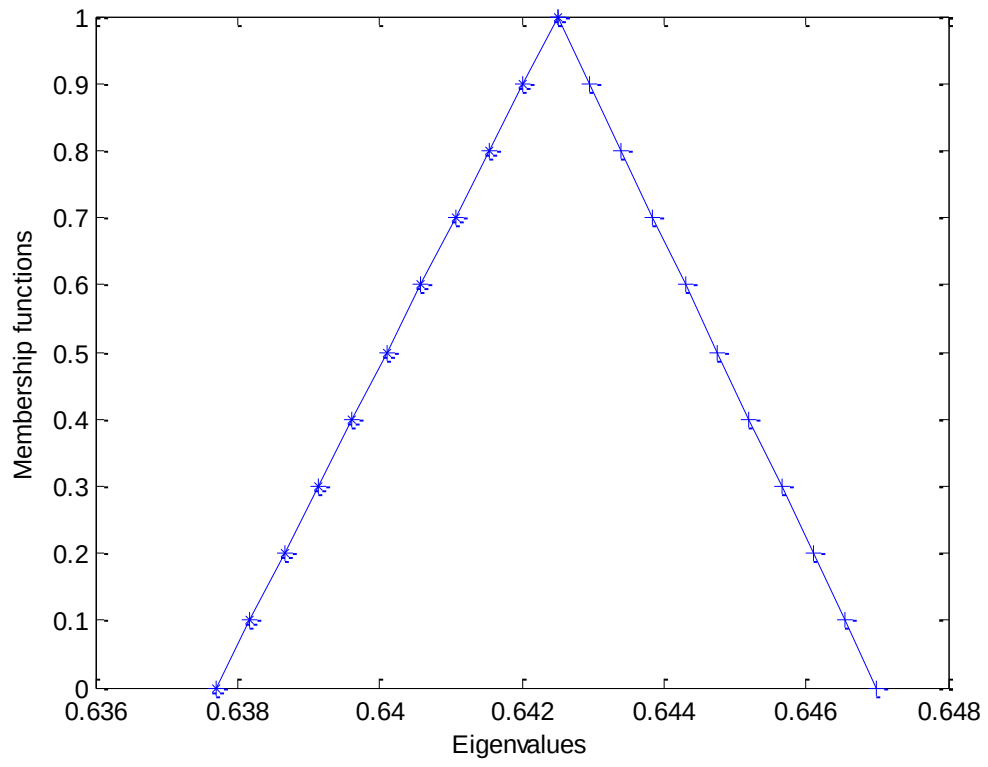

Figure 4 6 elements discretization

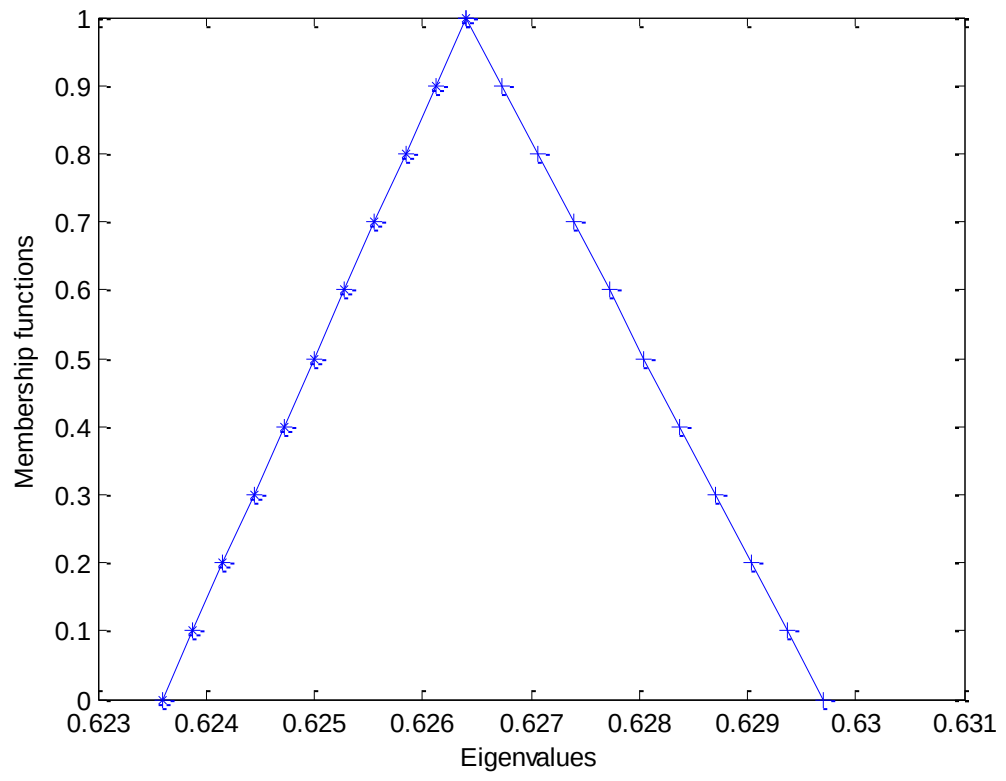

Figure 5 12 elements discretization

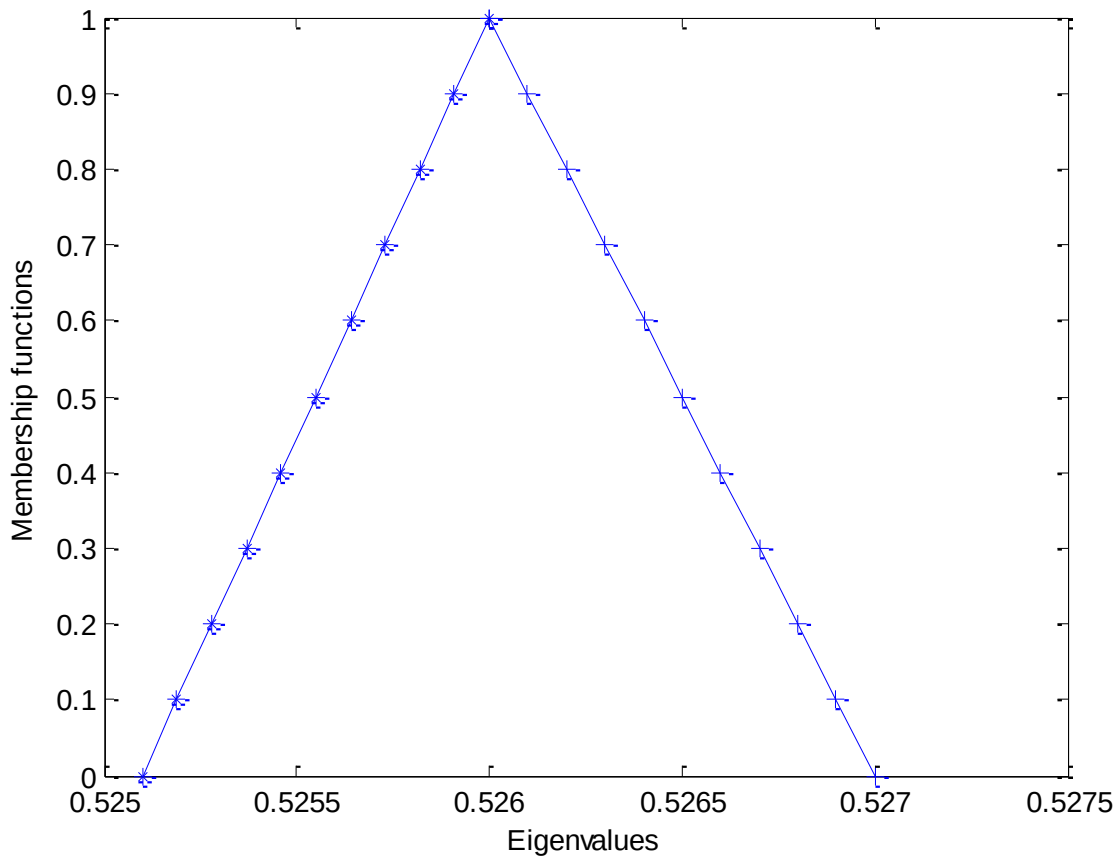

Figure 6 24 elements discretization

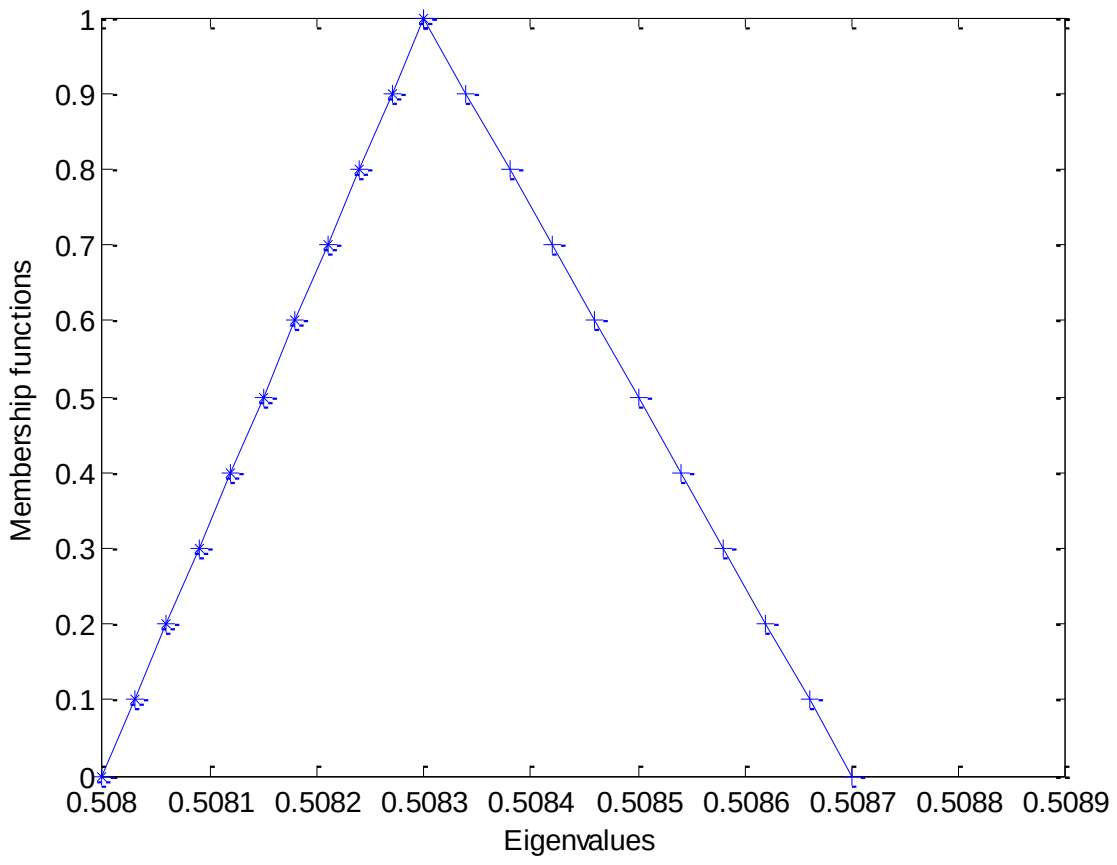

Figure 7 48 elements discretization

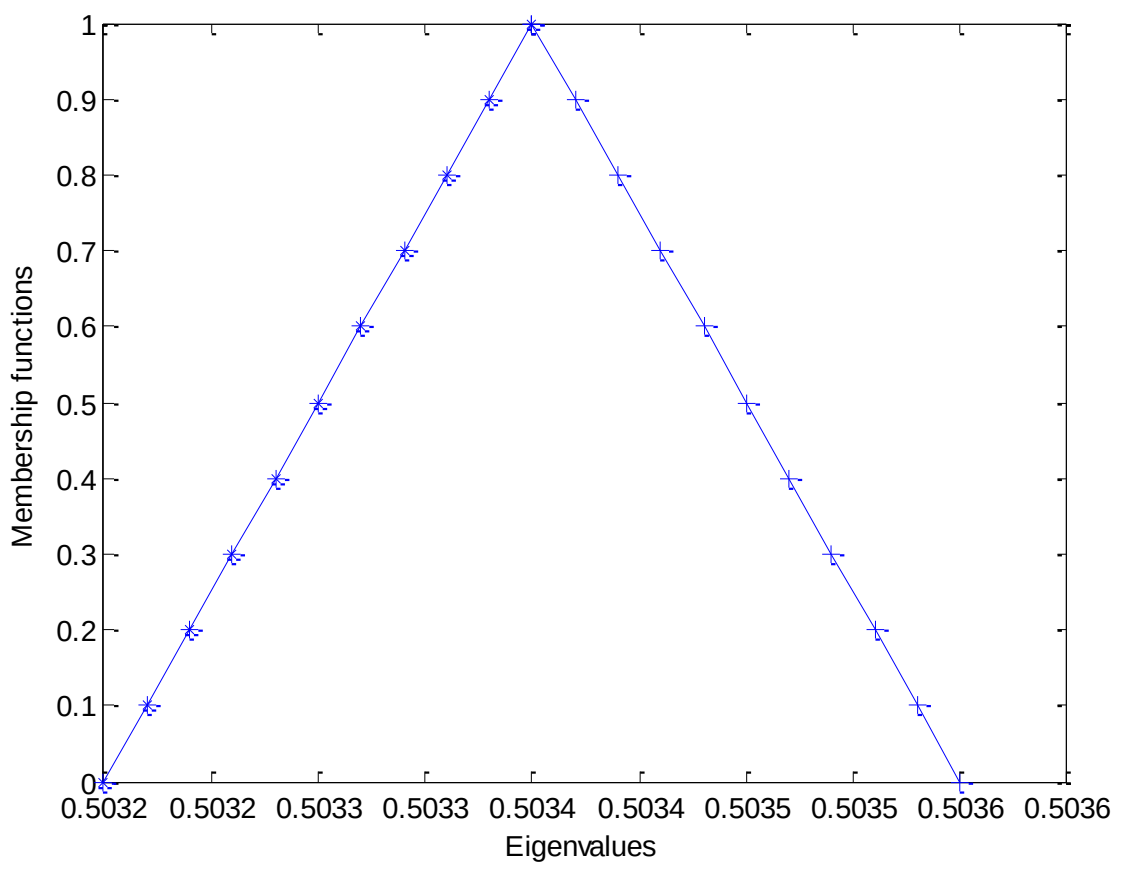

Figure 8 96 elements discretization

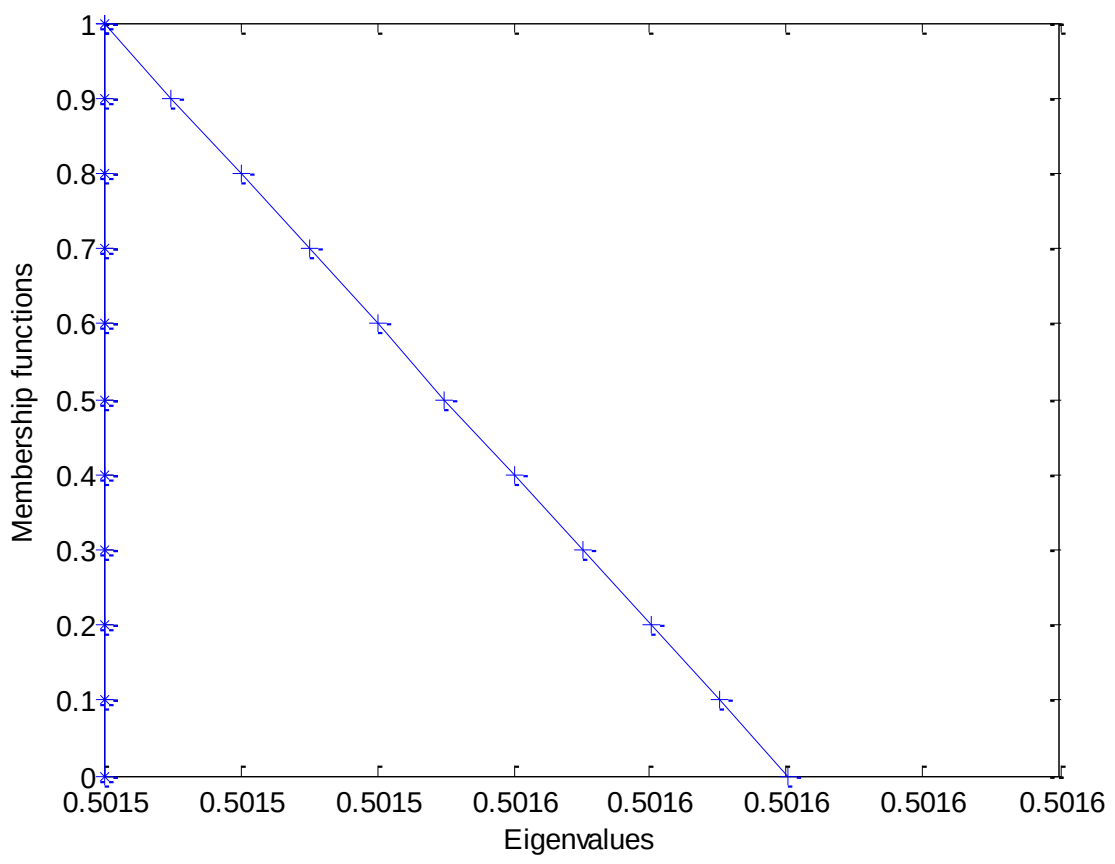

Figure 9 192 elements discretization

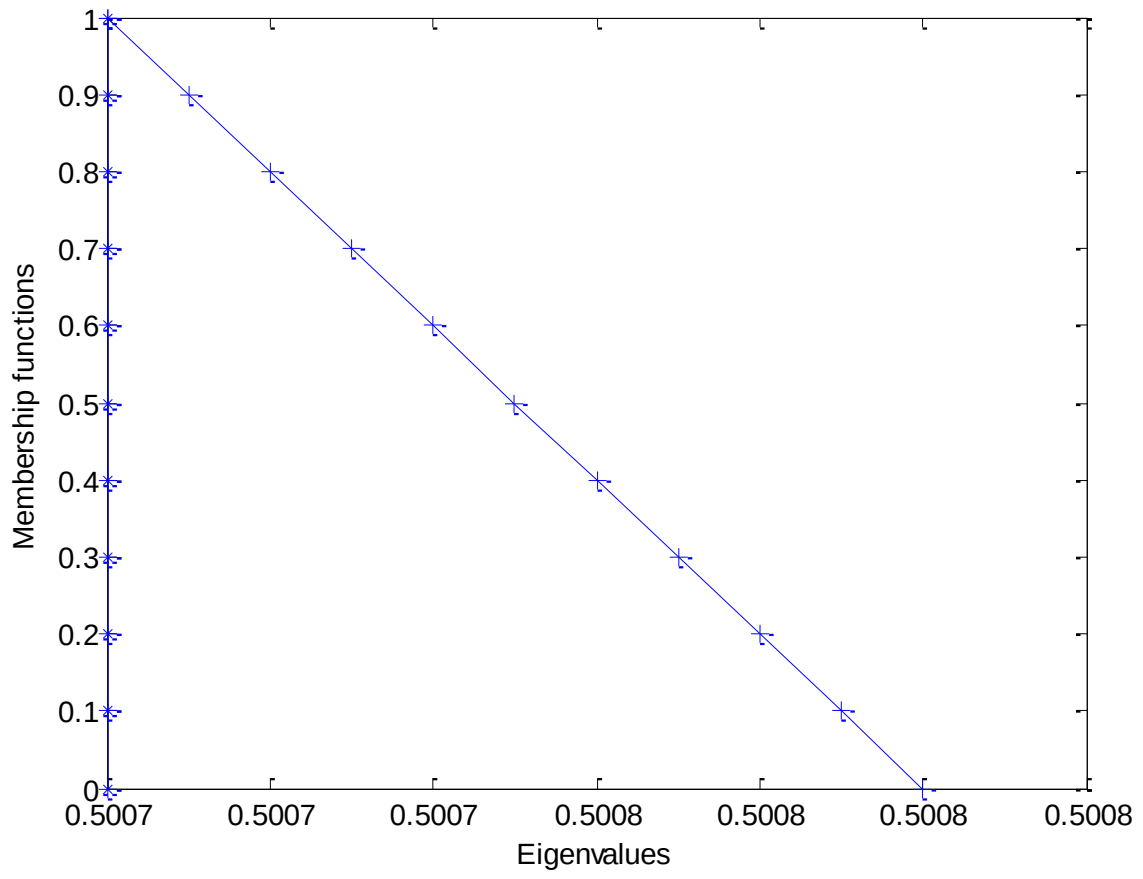

Figure 10 384 elements discretization

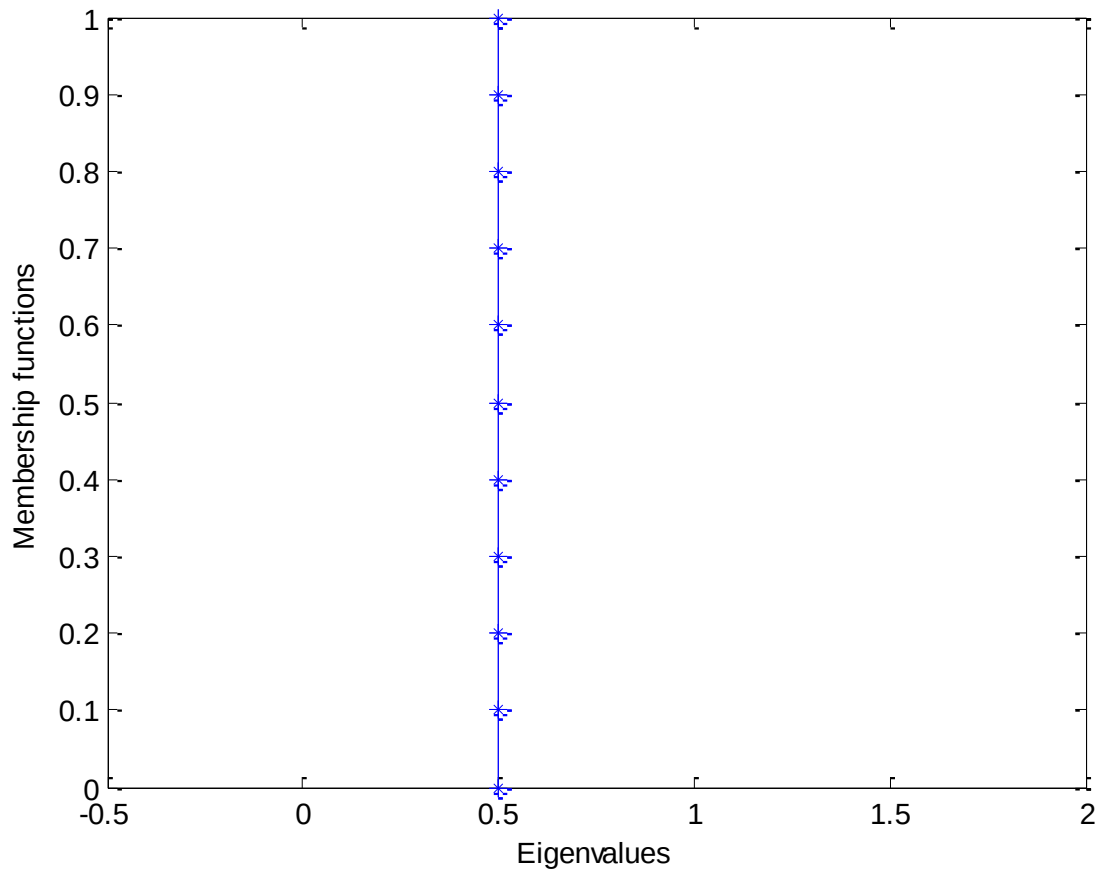

Figure 11 1536 elements discretization

The variation of both the crisp and fuzzy eigenvalues may be studied from Figure 12. Here a set of eigenvalues are given and the convergence is studied.

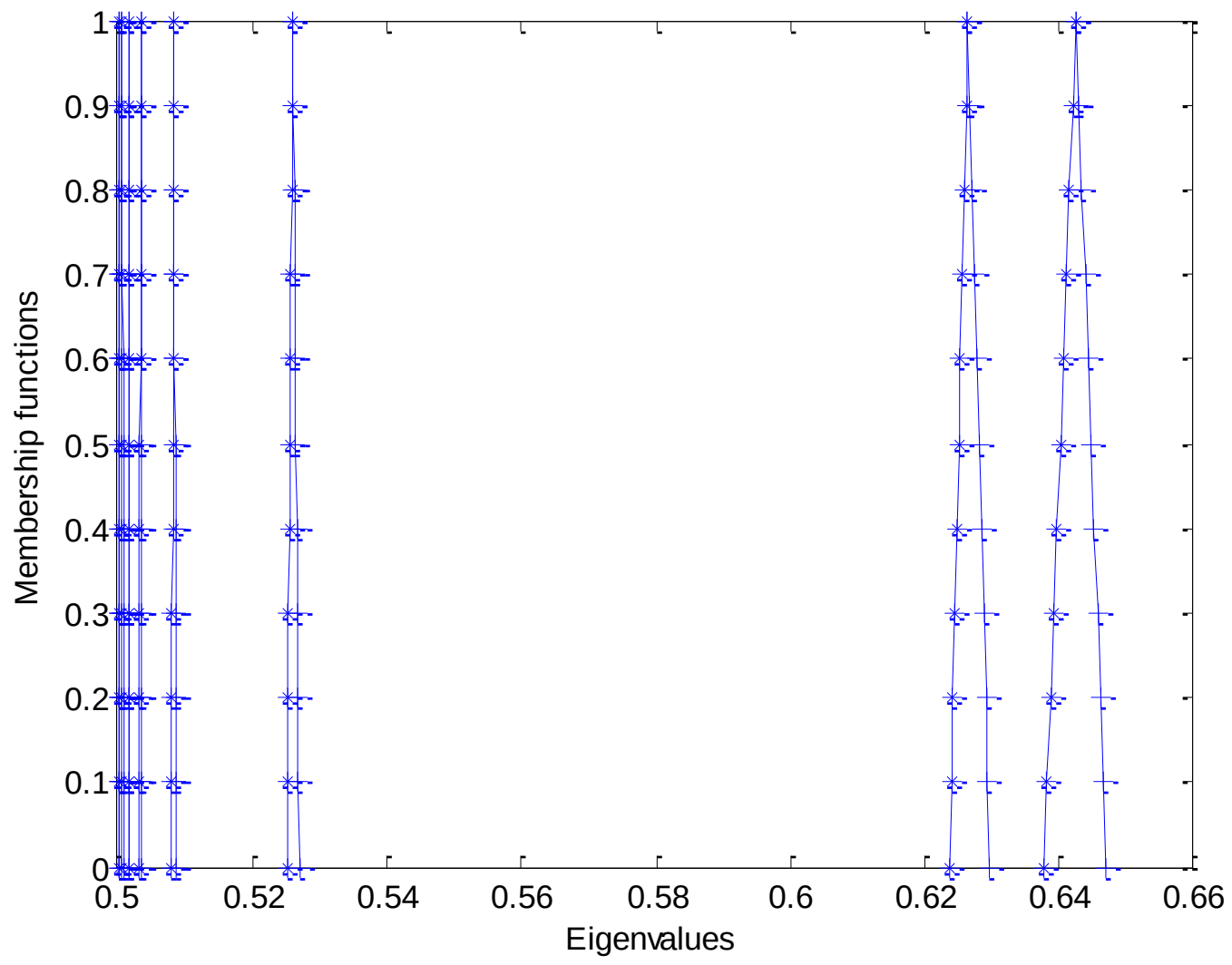

Figure 12 Triangular fuzzy membership functions for various discretizations of the domain

# 5. Discussion

When neutrons undergo diffusion in reactor then it suffers scattering collisions with the nuclei assumed to be initially stationary and they make different trajectory paths. Hence, to study the eigenvalue problem for corresponding one group neutron diffusion equation, we have considered triangular element discretization for one group neutron diffusion equation for a triangular bare reactor. Initially the eigenvalue problem is solved by classical finite element method for crisp parameters and then it is solved by proposed fuzzy finite element procedure.

Here triangular bare reactor is considered and neutron flux at centre of the triangular geometry is taken as zero. The geometry is discretized into number of triangular elements as given in Figure 3. So the neutron flux distributions are studied for other nodal points. Solving eigenvalue problem we get a set of eigenvalues for different number of element discretizations. It is seen that the eigenvalues are converging with respect to the increase in number of elements for discretized triangular bare reactor. The pattern of the convergence is presented in Table 2.

The diffusion and absorption coefficients are taken as the uncertain parameters. These uncertain parameters are considered as TFN to investigate the uncertain eigenvalues. Again the boundary condition for this uncertain case is taken same as that of the crisp case. As we increase the number of elements the uncertain eigenvalues gets converged. Further, uncertain width of eigenvalues decreases with increase in number of discretizations of the triangular bare homogeneous reactor.

From the obtained results it is observed that the shape of the TFN changes as we discretise the domain into more number of elements and these are shown in Figures 4 to 11. This variation of TFN occurs due to the left, right and centre values of the TFN. As we move on with the increase in number of discretization of the domain we get right angular shaped fuzzy number. The trends of the shapes are shown in Figures 8 to 11. In Figures 9 and 10 the left and centre values are obtained as same so we get left monotonically increasing function parallel to membership functions axis and width of the left bound from centre becomes zero. In Figure 11, the left, right and centre values approximately coincide and hence we get a straight line parallel to membership function axis. Here the variations of eigenvalues become constant with the change of membership functions. From Figure 12 it is seen that if we go on increasing the number of discretization of the said domain we get a series of uncertain eigenvalues and these triangular fuzzy eigenvalues converges to a constant value.

It may be noted that the reliability of the fuzzy results can be seen in the special cases viz. crisp and interval which are derived from the fuzzy values. As such three cases are reported with respect to the above.

**Case-1**

Here we have considered only left monotonic increasing functions of the resultant eigenvalues. The resulting eigenvalues vary with the value of membership functions. Assigning zero for the value of $\alpha$ we get the left bound of the uncertain fuzzy eigenvalues. Similarly if the value of $\alpha$ is taken as one then we get right bound of the left monotonic increasing functions which are the centre value of TFN.

**Case-2**

In this case only right monotonic decreasing functions of the resultant eigenvalues are considered. Resulting eigenvalues vary with the value of membership functions. Assigning zero for the value of $\alpha$ we get the right bound of the uncertain fuzzy eigenvalues. Similarly if the value of $\alpha$ is taken as one then we get left bound of the right monotonic decreasing functions which are the centre value of TFN.

**Case-3**

In this case let us consider the value of $\alpha$ is one for both the left and right monotonic functions. We observe that the resultant eigenvalues become same for both monotonic functions. If we consider TFN then we find that it is nothing but the centre value. Here we get an interval of eigenvalues where the membership functions are normalised.

# 6. Conclusion

The main purpose of this paper is to present an alternative non probabilistic method to manage various uncertain engineering and science problems. Here the traditional interval arithmetic is modified for the said problem and a simpler method is proposed to compute interval arithmetic. The idea of modified interval arithmetic is then extended for uncertain fuzzy numbers. As such uncertain parameters are taken as fuzzy and the fuzzy numbers are converted into interval using $\alpha$-cut techniques. These fuzzy numbers contain left monotonically increasing and right monotonically decreasing functions respectively.

In this paper we have considered one group of neutron diffusion equation for triangular bare reactor. The corresponding eigenvalues for one group of neutron diffusion equation for bare triangular homogeneous reactor is investigated. The presence of uncertain parameters makes the system uncertain and the uncertain eigenvalues are studied in detail. To handle uncertain system, the problem is modelled and presented by a modified fuzzy finite element method. In this study we found that proposed fuzzy finite element method for triangular element discretization gives a general procedure to solve the said problem. The generalization may be extended for various types of element discretizations. This method is found to be reliable by considering the convergence of uncertain width of the obtained results. It may be concluded that the fuzzy finite element method with the proposed interval computation is simpler to handle and also efficient. Hence it may be used as a tool to solve neutron diffusion problems for various other types of nuclear reactors.

# 7. Acknowledgement

The authors would like to thank BRNS (*Board of Research in Nuclear Sciences*), Department of Atomic Energy, (DAE), Govt. of India for providing fund to do this work.